\documentclass[%
 reprint,
 superscriptaddress,
 showpacs,amsmath,amssymb,
 aps,prl,
floatfix
]{revtex4-1}
\usepackage{graphicx}
\usepackage{dcolumn}
\usepackage{bm}
\usepackage[mathlines]{lineno}
\usepackage{color,soul}

\begin{document}

\preprint{APS/123-QED}

\title{Extended Search for the Invisible Axion with the Axion Dark Matter Experiment}
\author{T. Braine}
  \affiliation{University of Washington, Seattle, WA 98195, USA}
\author{R. Cervantes}
  \affiliation{University of Washington, Seattle, WA 98195, USA}
 \author{N. Crisosto}
   \affiliation{University of Washington, Seattle, WA 98195, USA}
\author{N. Du}%
  \email[Correspondence to: ]{ndu@uw.edu}
  \affiliation{University of Washington, Seattle, WA 98195, USA}
\author{S. Kimes}
  \affiliation{University of Washington, Seattle, WA 98195, USA}
 \author{L. J Rosenberg}%
  \affiliation{University of Washington, Seattle, WA 98195, USA}
  \author{G. Rybka}%
  \affiliation{University of Washington, Seattle, WA 98195, USA}
    \author{J. Yang}%
  \affiliation{University of Washington, Seattle, WA 98195, USA}
  
%
\author{D. Bowring}
  \affiliation{Fermi National Accelerator Laboratory, Batavia IL 60510, USA}
\author{A. S. Chou} 
  \affiliation{Fermi National Accelerator Laboratory, Batavia IL 60510, USA}
    \author{R. Khatiwada}
  \affiliation{Fermi National Accelerator Laboratory, Batavia IL 60510, USA}
\author{A. Sonnenschein} 
  \affiliation{Fermi National Accelerator Laboratory, Batavia IL 60510, USA}
  \author{W. Wester} 
  \affiliation{Fermi National Accelerator Laboratory, Batavia IL 60510, USA}

\author{G. Carosi}
\affiliation{Lawrence Livermore National Laboratory, Livermore, CA 94550, USA}
\author{N. Woollett}
\affiliation{Lawrence Livermore National Laboratory, Livermore, CA 94550, USA}

\author{L. D. Duffy}
  \affiliation{Los Alamos National Laboratory, Los Alamos, CA 87545, USA}

\author{R. Bradley}
  \affiliation{National Radio Astronomy Observatory, Charlottesville, Virginia 22903, USA}

\author{C. Boutan}
  \affiliation{Pacific Northwest National Laboratory, Richland, WA 99354, USA}
\author{M. Jones}
  \affiliation{Pacific Northwest National Laboratory, Richland, WA 99354, USA}
\author{B. H. LaRoque}
  \affiliation{Pacific Northwest National Laboratory, Richland, WA 99354, USA}
\author{N. S. Oblath}
  \affiliation{Pacific Northwest National Laboratory, Richland, WA 99354, USA}
\author{M. S. Taubman}
  \affiliation{Pacific Northwest National Laboratory, Richland, WA 99354, USA}

\author{J. Clarke}
  \affiliation{University of California, Berkeley, CA 94720, USA}
\author{A. Dove}
  \affiliation{University of California, Berkeley, CA 94720, USA}
\author{A. Eddins}
  \affiliation{University of California, Berkeley, CA 94720, USA}
\author{S. R. O'Kelley}
  \affiliation{University of California, Berkeley, CA 94720, USA}
\author{S. Nawaz}
  \affiliation{University of California, Berkeley, CA 94720, USA}
\author{I. Siddiqi}
  \affiliation{University of California, Berkeley, CA 94720, USA}
\author{N. Stevenson}
  \affiliation{University of California, Berkeley, CA 94720, USA}

\author{A. Agrawal}
\affiliation{University of Chicago, IL 60637, USA}
\author{A. V. Dixit}
\affiliation{University of Chicago, IL 60637, USA}
 
\author{J.~R.~Gleason}
  \affiliation{University of Florida, Gainesville, FL 32611, USA}
\author{S. Jois}
  \affiliation{University of Florida, Gainesville, FL 32611, USA}
 \author{P. Sikivie}
  \affiliation{University of Florida, Gainesville, FL 32611, USA}
\author{N. S. Sullivan}
  \affiliation{University of Florida, Gainesville, FL 32611, USA}
\author{D. B. Tanner}
  \affiliation{University of Florida, Gainesville, FL 32611, USA}
  
\author{E. Lentz}
  \affiliation{University of G\"{o}ttingen, G\"{o}ttingen, Germany}  
  
\author{E. J. Daw}
  \affiliation{University of Sheffield, Sheffield, UK}
  
\author{J. H. Buckley}
  \affiliation{Washington University, St. Louis, MO 63130, USA}
\author{P. M. Harrington}
  \affiliation{Washington University, St. Louis, MO 63130, USA}
\author{E. A. Henriksen}
  \affiliation{Washington University, St. Louis, MO 63130, USA} 
\author{K. W. Murch}
  \affiliation{Washington University, St. Louis, MO 63130, USA}

\collaboration{ADMX Collaboration}\noaffiliation

\date{\today}

\begin{abstract}
This paper reports on a cavity haloscope search for dark matter axions in the galactic halo in the mass range $2.81$--$3.31$ ${\mu}eV$. This search excludes the full range of axion-photon coupling values predicted in benchmark models of the invisible axion that solve the strong CP problem of quantum chromodynamics, and marks the first time a haloscope search has been able to search for axions at mode crossings using an alternate cavity configuration. Unprecedented sensitivity in this higher mass range is achieved by deploying an ultra low-noise Josephson parametric amplifier as the first-stage signal amplifier. 
\end{abstract}

\maketitle

 
Axions are a hypothesized particle that emerged as a result of the Peccei-Quinn solution to the strong CP problem  \cite{Peccei:1977hh,Weinberg:1977ma,Wilczek:1977pj}. In addition, axions are a leading dark-matter candidate that could explain $100\%$ of the dark-matter in the Universe \cite{Planck, Abbott:1982af,DINE1983137,Preskill:1982cy,PhysRevLett.50.925}. There are a number of mechanisms for the production of dark-matter axions in the early Universe \cite{Abbott:1982af,DINE1983137,Sikivie:2006ni,PhysRevD.78.083507}. For the case where $U_{\rm PQ}(1)$ becomes spontaneously broken after inflation, cosmological constraints suggest an axion mass on the scale of $1$~${\mu}eV$ or greater \cite{Bonati2016,PhysRevD.92.034507,Borsanyi2016,PhysRevLett.118.071802,PhysRevD.96.095001,PETRECZKY2016498}. Two benchmark models for the axion are the Kim-Shifman-Vainshtein-Zakharov (KSVZ) \cite{Kim:1979if,Shifman:1979if} and Dine-Fischler-Srednicki-Zhitnitsky (DFSZ) \cite{Dine:1981rt,Zhitnitsky:1980tq} models. Of the two, the DFSZ model is especially compelling because of its grand unification properties \cite{Dine:1981rt}.  

The Axion Dark Matter eXperiment (ADMX) searches for dark-matter axions using an axion haloscope \cite{Sikivie:1983ip}, which consists of a microwave resonant cavity inside a magnetic field. In the presence of an external magnetic field, axions inside the cavity can convert to photons with frequency $f=E/h$, where $E$ is the total energy of the axion, including the axion rest mass energy, plus a small kinetic energy contribution. The power expected from the conversion of an axion into microwave photons in the ADMX experiment is extremely low, $\mathcal{O}(10^{-23}~W)$, requiring the use of a dilution refrigerator and an ultra low-noise microwave receiver to detect the photons. 

In limits set in a previous paper, ADMX became the only axion haloscope to achieve sensitivity to both benchmark axion models for axion masses between $2.66$ and $2.81$ ${\mu}eV$ \cite{PhysRevLett.120.151301}. This paper reports on recent operations which extend the search for axions at DFSZ sensitivity to $2.66$--$3.31$ ${\mu}eV$.

The ADMX experiment consists of a $136$-liter cylindrical copper-plated cavity placed in a $7.6$-T field produced by a superconducting solenoid magnet. The magnet and cavity configuration are similar to the configuration described in Ref.~\cite{Asztalos:2009yp,ASZTALOS201139}. A magnetic field-free region above the cavity is maintained by a counter-wound bucking magnet above the cavity. Field sensitive receiver components, such as a Josephson parametric amplifier (JPA) and circulators, are located there, and the JPA is protected by additional passive magnetic shielding.

The resonant frequency of the cavity is set by two copper tuning rods that run parallel to the axis of the cavity and can be positioned between near the center of the cavity and the walls of the cavity. Cryogenic gearboxes connect the tuning rods to room-temperature stepper motors, which  tune the cavity during operations. When the frequency of the cavity is tuned to the same frequency as the photon produced from the axion, the expected power deposited into the cavity is \cite{Sikivie1985}

\begin{equation}
\begin{split}
    P_{\mathrm{axion}}=2.2\cdot10^{-23}~\text{W} (\frac{V}{136~\text{L}})(\frac{B}{7.6~\text{T}})^2(\frac{C}{0.4})\\
    \cdot(\frac{g_{\gamma}}{0.36})^2(\frac{\rho_a}{0.45~\text{GeV cm}^{-3}})(\frac{f}{740~\text{MHz}})(\frac{Q}{30000}).
\end{split}
\label{Eq:Axion_Power}
\end{equation}

\noindent Here $V$ is the volume of the cavity, $B$ is the magnitude of the external magnetic field,  $g_{\gamma}$ is the model dependent axion-photon coupling, which has a value of $-0.97$ ($0.36$) for KSVZ (DFSZ) axions, $\rho_a$ is the local dark-matter density, $f$ is the frequency of the photon, $Q$ is the loaded quality factor of the cavity, and $C$ is the form factor of the cavity.

The form factor is defined as the amount of overlap between the electric field of the cavity mode and the external magnetic field generated by the solenoid \cite{Sikivie1985}. In the case of the ADMX cavity, the optimal form factor is observed with the TM$_{010}$-like lowest order tunable mode. Over the mass range explored in this paper, the average form factor is $0.4$. 

Several mode crossings between the TM$_{010}$ mode frequency and weakly tuning TE or TEM modes occurred during operations, causing a significant drop in the form factor due to mode mixing. In order to fill in the mode crossings, alternative rod configurations were used which shifted the static mode by several MHz, thereby reducing the form factor in the region of the mode crossing by only a moderate amount. 


The signal-to-noise ratio for power within the experiment is set by the Dicke radiometer equation \cite{doi:10.1063/1.1770483}
\begin{equation}
    \frac{S}{N}=\frac{P_\mathrm{axion}}{k_BT_\mathrm{sys}}\sqrt{\frac{t}{b}},
\end{equation}
where $T_\mathrm{sys}$ is the system noise temperature, equal to the combined physical temperature of the cavity and the noise temperature of the receiver chain, $t$ is the integration time, and $b$ is the detection bandwidth. 

To reduce the noise temperature, the cavity and JPA are cooled with a Janis Research dilution refrigerator. The mixing chamber of the dilution refrigerator is mounted to the top of the cavity, and its high cooling power enables ongoing operation of the cavity and JPA at temperatures on the order of $100$ mK. 

Ruthenium oxide temperature sensors measured the cavity temperature to be typically $130$ mK, and the temperature of the receiver amplifier package was $230$ mK. The higher temperature in the region of the JPA was due to weak thermal contact between the cryogenic receiver package and the 4-K liquid helium reservoir surrounding the magnetic field-free region. 

\begin{figure}
    \centering
    \includegraphics[width=\linewidth]{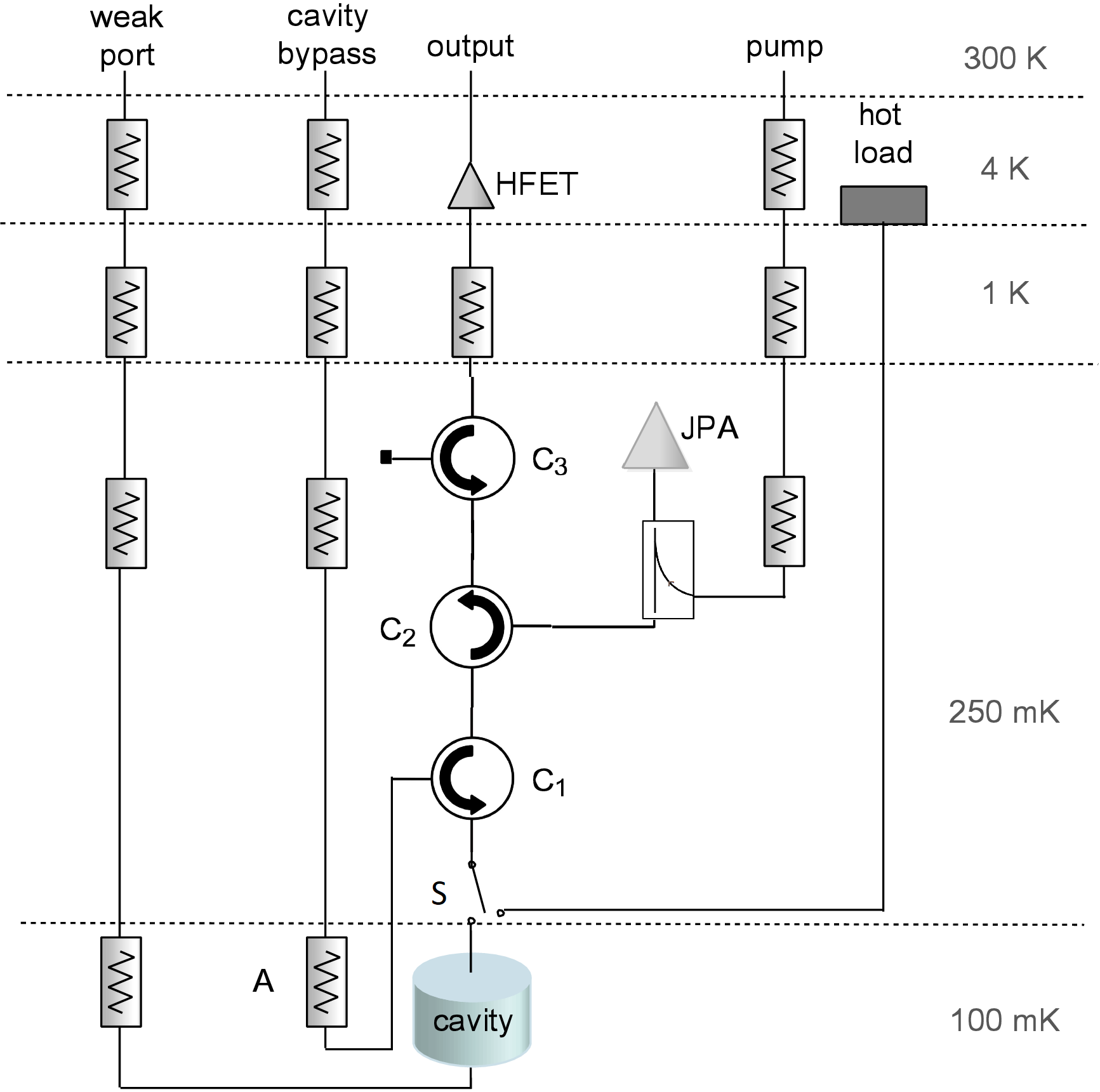}
    \caption{The ADMX cryogenic receiver chain. C$_1$, C$_2$, and C$_3$ are circulators. JPA is the Josephson parametric amplifier and HFET is a heterostructure field-effect transistor amplifier. Both are used to amplify power from the cavity. Power is transmitted into the weak port and cavity bypass lines for transmission and reflection measurements from the cavity, respectively. The pump line is used to supply a pump tone into the JPA. Switch S can be toggled between the cavity and the hot load for noise calibration measurements.}
    \label{fig:rf_layout}
\end{figure}

\begin{figure*}[htb!]
    \centering
    \includegraphics[width=0.8\textwidth]{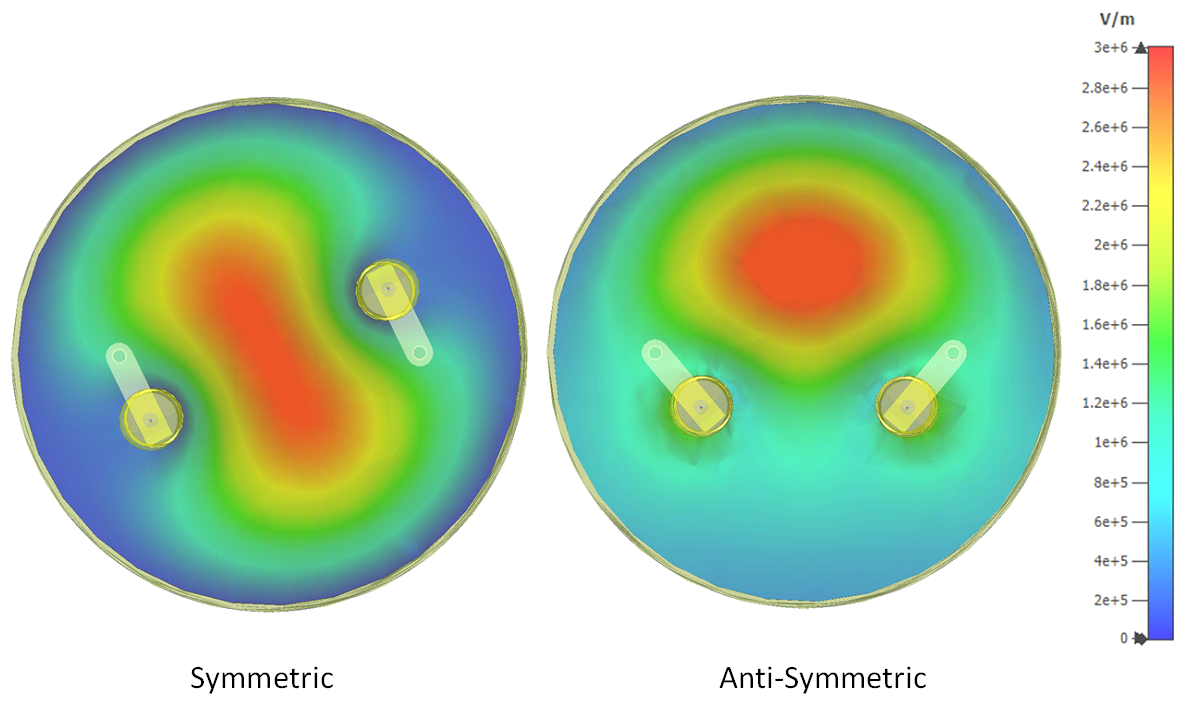}
    \caption{A top-down view of the ADMX cavity. The resonant frequency of the cavity is set by the position of two tuning rods. During initial data-taking, the rods are moved in a symmetric configuration (left). To scan over mode crossings, an anti-symmetric configuration is used (right). The frequency of the TM$_{010}$ mode is the same in both configurations shown. The colors indicate the magnitude of the electric field component along the axis of the cavity.}   
    \label{fig:rod_configuration}        
\end{figure*}

Data were collected between January and October 2018, and mark the first time the ADMX experiment collected data using a JPA. During standard data-taking operations, small steps in tuning rod position were taken to adjust the resonant frequency of the cavity. The cavity resonant frequency and $Q$ were measured by injecting power through a weakly coupled antenna on the cavity and measuring the transmission of the cavity. A $50$-kHz wide power spectrum, centered on the cavity resonant frequency, was constructed by integrating the voltage time series from the cavity with a digitizer for $100$ seconds. Periodically during data taking, the coupling of the antenna to the cavity mode was determined with an effective $S_{11}$ reflection measurement to measure the power reflected from the cavity antenna. If the coupling between the cavity and receiver fell below $5$~dB, the antenna position was adjusted. The JPA was operated in a phase insensitive mode by pumping with a microwave tone $375$ kHz detuned from the cavity resonance.

During data taking, synthetically generated axion signals were occasionally injected into the weakly coupled antenna. The frequencies of these synthetic axion injections were blinded to the group searching for axion candidates in the data. 

The system noise temperature was measured in a two-step process. The noise contribution from the HFET amplifiers and warm receiver chain were calibrated from a Y-factor measurement, followed by a signal-to-noise improvement measurement to determine the noise contribution from the JPA \cite{wilson2011techniques}. For the Y-factor measurement, the JPA was unpowered, where it operated as a passive mirror, and switch S was toggled between the cavity and a dedicated hot load.  As shown in Fig.~\ref{fig:rf_layout}, when toggled to the cavity, thermal photons from attenuator A were transmitted through circulator C$_1$, and reflected off the cavity. Attenuator A was thermally sunk to the cavity and could be varied between $100$ mK and $1$ K.  When toggled to the dedicated load, the thermal background was dominated by the hot load, which was heated between $4$ to $6$ K. The power from the hot load was attenuated through the quantum amplifier package, which could be varied between $200$ and $500$ mK, independent of the cavity and dedicated load. These measurements were repeated four times during the data-taking period.  The measurements were combined and fitted to a model of the RF system, which yielded a downstream (of all components beyond the JPA electronics space) noise temperature of $11.3\pm0.1$ K over the band of interest which was within expectations for the HFET amplifier operating in a high magnetic field \cite{doi:10.1063/1.366000}. These measurements also identified the attenuation between the cavity and the JPA  to be $1.52\pm0.02$ dB, consistent with the expected loss in the circulators, cables, and JPA below 780 MHz.  This attenuation was larger above 780 MHz, as expected from the circulator specifications, and incorporated into the calculation of axion sensitivity.

The signal-to-noise ratio improvement (SNRI) measurement consisted of measuring the increase in the digitized power from the cavity and the gain of the receiver with and without the JPA pump tone active. The gain of the JPA multiplied by the ratio of the power spectral density measurements yielded the ratio of the total system noise temperature to the noise temperature of the HFET and downstream components, such that 

\begin{equation}
    T_\mathrm{sys}=\frac{G_\mathrm{off}}{G_\mathrm{on}}\frac{P_\mathrm{on}}{P_\mathrm{off}}T_\mathrm{hfet}=\frac{T_\mathrm{hfet}}{SNRI},
    \label{eqn:snr_improvement}
\end{equation}
where $G$ was the transfer function at the desired frequency, and $P$ was the power at the desired frequency. A typical SNRI measured was $15.5$ dB, which yielded a typical system noise temperature of $350$ mK.

The system noise was monitored by SNRI measurements roughly every 10 minutes and at eight different nearby JPA-bias and pump-power combinations, which were then used to update the bias current and pump power for the optimal JPA SNRI. 

The initial axion search was performed with the tuning rods in a symmetric configuration and in that configuration eight mode crossings were observed.  These crossings were identified with simulations and confirmed with wide-span $S_{21}$ transmission measurements. During initial data taking, the frequency ranges near these mode crossings were skipped over because of their poor form factors.

To fill in mode crossings, an anti-symmetric rod configuration was used. In this rod configuration, many weakly tuning modes were shifted by several MHz, shifting the positions of the corresponding mode crossings. Examples of both rod configurations are shown in Fig.~\ref{fig:rod_configuration}. At frequency ranges previously covered by mode crossings, the form factor was boosted to about $0.35$, which was sufficient for axion searches; however, because the unloaded $Q$-values in the anti-symmetric configuration were increased by a factor of $1.5$ over the symmetric configuration, the overall sensitivity to axions remained similar. Three mode crossings were covered by this method. This represents the first time in a haloscope search that mode-crossings were filled in using an alternative tuning rod configuration. 

The five remaining mode crossings were either too wide or the interfering mode could not be shifted, so that we were unable to obtain sufficient sensitivity to the axion to set limits at those frequencies. Those mode crossings and their corresponding widths are listed in Table~\ref{tab:regions}.

\begin{table}
    \centering
    \begin{tabular}{|c|c|l|}
    \hline
    Mode Crossing Frequency (MHz) & Width (MHz) \\
    \hline
      704.659 & 0.350  \\
    715.064 & 0.140  \\
    717.025 & 0.140 \\
    726.624 & 0.701  \\
    753.844 & 12.682 \\
      \hline
    \end{tabular}
    \caption{Mode crossing locations where an exclusion limit could not be set.}
    \label{tab:regions}
\end{table}

The analysis procedures followed those discussed in Ref.~\cite{Brubaker:2017rna}. The goal of the analysis was to average the individual spectra into a single grand spectrum to increase the signal-to-noise ratio of possible axion signals, and then search for candidate axion signals. First, the receiver shape was removed with a fit to a six-order Pad\'e approximation. The power was scaled to the system noise temperature and weighted by the cavity $Q$ to convert the spectrum into a measurement of the power in excess of the noise. This spectrum was then filtered by a convolution with one of two different axion signal shapes: a boosted Maxwell-Boltzmann line shape, predicted from the standard halo model for axion dark-matter, as described in \cite{PhysRevD.42.3572}, with a local density of $0.45$ $\mathrm{GeV/cm}^3$, or a line shape derived from N-body simulations described in \cite{0004-637X-845-2-121}, with a local density of $0.63$ $\mathrm{GeV/cm}^3$. 

\begin{figure}[htb!]
    \centering
    \includegraphics[width=\linewidth]{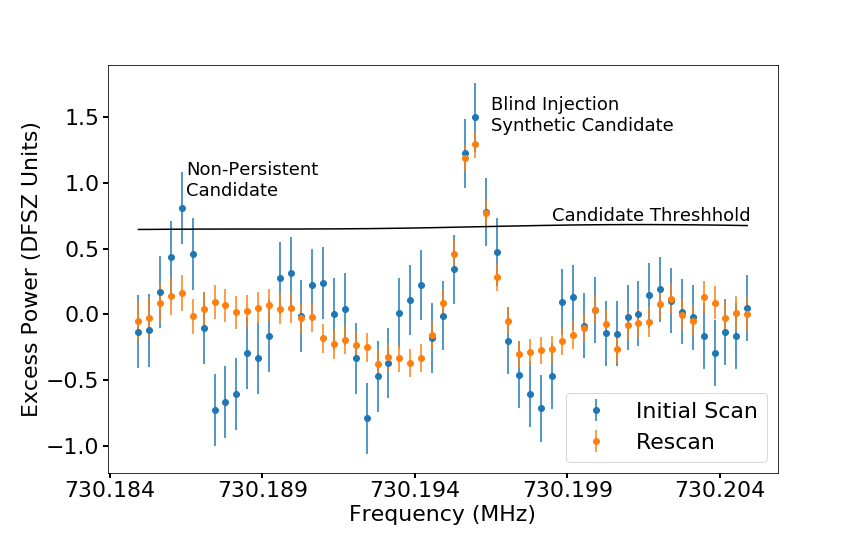}
    \caption{An example of combined power spectra after a Maxwell-Boltzmann shape filter, with blue indicating the initial scan data and orange indicating data taken during a rescan with roughly 4 times more integration time.  The prominent peak centered at $730.195$ MHz corresponds to a blind signal injection identified in the analysis that persisted after a rescan; the small peak to the left at $730.186$ MHz was a candidate that did not persist in the rescan. Because of a mismatch between the receiver spectral shape and the axion signal, the power at frequencies surrounding the candidate is suppressed by the receiver spectral background removal. This can be seen in the frequency background surrounding the $730.195$ MHz candidate in the rescan (orange) data.}
    \label{fig:synthetic_candidate_example}
\end{figure}

After a frequency range had been scanned with a more than sufficient signal-to-noise ratio to exclude DFSZ axions in the null case, ``candidate" axion signals were identified. Frequencies with upward fluctuations in power exceeding 3$\sigma$ or that could not exclude the DFSZ axion coupling strength were deemed ``candidate" axion signals, requiring rescanning and further analysis. At frequencies with no statistically significant power excess, upper limits were placed on the axion-photon coupling using the measured power and uncertainty at that frequency.

Following the initial search, candidate frequencies were rescanned with significantly longer integration time to improve the expected signal-to-noise for a possible axion.  If the power at the candidate frequency did not persist and a DFSZ axion signal could be excluded, the candidate was determined to be transient. Frequencies that persisted past the second rescan were subjected to additional individual-candidate checks.


\begin{figure*}[htb!]
    \centering
    \includegraphics[width=0.9\linewidth]{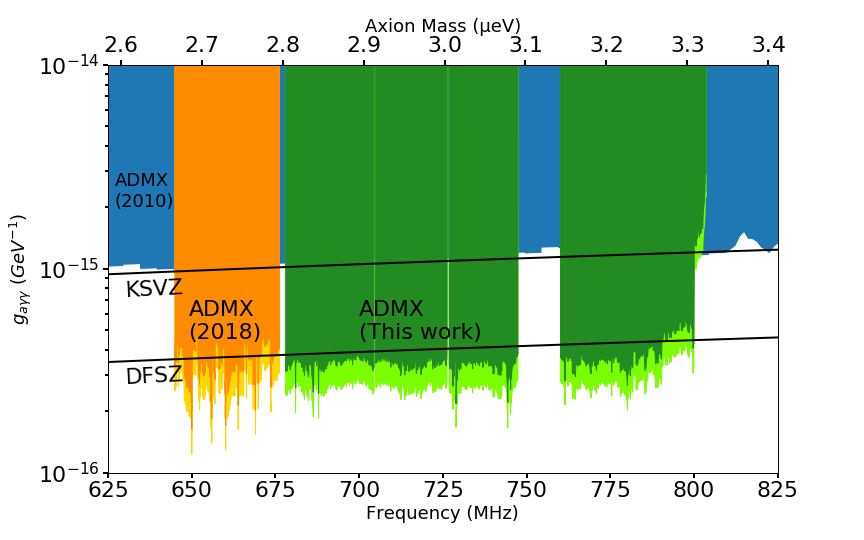}
    \caption{$90\%$ confidence exclusion on axion-photon coupling as a function of axion mass for the Maxwell-Boltzmann (MB) dark-matter model and N-body model.  \textit{Blue} Previous limits reported in \cite{PhysRevD.64.092003}. \textit{Orange} Previous limits reported in \cite{PhysRevLett.120.151301}. \textit{Green} Limits from this work. Darker shades indicates limits set for the MB model \cite{PhysRevD.42.3572} and the lighter shade indicates limits set for the N-body model \cite{0004-637X-845-2-121}.}
    \label{fig:limit_plot}
\end{figure*}

The first step of individual-candidate analysis was to check whether candidates were external radio signals detected within the experiment by measuring whether the signal power was maximized when the cavity frequency was tuned to the signal frequency. Axion signals would develop within the resonant cavity, so that the power from an axion signal would increase as the cavity was tuned to match the signal frequency. On the other hand, external interference would be picked up by components further along in the receiver chain, so that the signal power would be independent of the cavity frequency.

To test the detection efficiency of the analysis procedure, $20{,}000$ software injected signals were injected with powers between those expected for DFSZ and KSVZ axions, and the injected powers were compared to the powers detected by the analysis.  It was found that the detected power was suppressed by a factor of $0.82$ due to signal removal from the receiver spectral fit. This effect is accounted for in our reported limits.

Over the course of the axion search, ADMX searched for axions from $680$--$790$ MHz. Within this range, three persistent signals were observed, at $686.310$, $730.195$, and $780.255$ MHz. The first and last signals ($686.310$ and $780.255$ MHz) did not maximize on-resonance, indicating they were due to external radio interference and therefore could be excluded as axions.  

The signal at $730.195$ MHz (Fig.~\ref{fig:synthetic_candidate_example}) maximized on resonance and was consistent in power and linewidth to the signal expected from a DFSZ axion. This result triggered a decision to ramp the magnetic field down to determine whether the power of the signal would scale as $B^2$, in a manner consistent with an axion signal. Before the procedure was initiated, the candidate was revealed to be a synthetic axion signal. Instead the synthetic injection was disabled and the region around the candidate was rescanned. No signal appeared at $730.195$ MHz, and thus all candidate axion signals were excluded. We concluded that either the axion was not within the explored range, that the axion dark-matter density is a small fraction of the halo density, or that the axion-photon coupling constant is significantly below the prediction for DFSZ.

Given the absence of axion-like signals, a $90\%$ upper confidence limit was set on the axion-photon coupling over the scanned mass range.  Due to the loss of sensitivity at mode crossings, we do not report limits over some regions; a detailed list is given in Table~\ref{tab:regions}. For models where axions make up 100\% of dark-matter, these limits exclude DFSZ axion-photon couplings between $2.66$ and $3.31$ ${\mu}eV$ for both isothermal sphere halo models and N-body simulations (Fig.~\ref{fig:limit_plot}). These results represent a factor of four increase in mass coverage over those reported in \cite{PhysRevLett.120.151301}.


ADMX will utilize a similar cavity with larger tuning rods and improved thermalization between the dilution refrigerator and quantum amplifier package to continue to search dark-matter axions at higher masses with increased sensitivity. These future searches, built on current research and development \cite{thesisBoutan,PhysRevLett.121.261302}, will probe even more deeply into the well-motivated yet unexplored axion parameter space. A discovery could be made at any moment.
 
This work was supported by the U.S. Department of Energy through Grants Nos. DE-SC0009723, DE-SC0010296, DE-SC0010280, DE-SC0010280, DE-FG02-97ER41029, DE-FG02-96ER40956, DE-AC52-07NA27344, and DE-C03-76SF00098. This manuscript has been authored by Fermi Research Alliance, LLC under Contract No. DE-AC02-07CH11359 with the U.S. Department of Energy, Office of Science, Office of High Energy Physics. Additional support was provided by the Heising-Simons Foundation and by the LDRD offices of the Lawrence Livermore and Pacific Northwest National Laboratories. LLNL Release Number: LLNL-JRNL-763299.

\bibliographystyle{apsrev4-1}
\bibliography{2019admx_main}

\end{document}